\documentclass[conference]{IEEEtran}
\IEEEoverridecommandlockouts
\usepackage{cite}
\usepackage{amsmath,amssymb,amsfonts}
\usepackage{algorithmic}
\usepackage{graphicx}
\usepackage{textcomp}
\usepackage{xcolor}
\usepackage{stfloats}
\usepackage{float}

\ifCLASSOPTIONcompsoc
  \usepackage[caption=false,font=normalsize,labelfont=sf,textfont=sf]{subfig}
\else
  \usepackage[caption=false,font=footnotesize]{subfig}

\def\BibTeX{{\rm B\kern-.05em{\sc i\kern-.025em b}\kern-.08em
    T\kern-.1667em\lower.7ex\hbox{E}\kern-.125emX}}
\begin{document}

\title{Swin Transformer-Based CSI Feedback for\\ Massive MIMO
\thanks{This work is supported by the Natural Science Foundation of China (62122012, 62221001); the Beijing Natural Science Foundation (L202019, L211012); the Fundamental Research Funds for the Central Universities (2022JBQY004); ZTE Industry-University-Institute Cooperation Funds under Grant No. HC-CN-20221208006.
}
}

\author{\IEEEauthorblockN{Jiaming Cheng$^{1,2}$, Wei Chen$^{1,2,3}$, Jialong Xu$^{1,2}$, Yiran Guo$^{1,2}$, Lun Li$^{5,6}$, Bo Ai$^{1,2,4}$}
\IEEEauthorblockA{
$^{1}$ State Key Laboratory of Advanced Rail Autonomous Operation, Beijing Jiaotong University, China \\
$^{2}$ School of Electronic and Information Engineering, Beijing Jiaotong University, Beijing, China \\
$^{3}$ Frontiers Science Center for Smart High-speed Railway System \\
$^{4}$ Key Laboratory of Railway Industry of Broadband Mobile Information Communications \\
$^{5}$ State Key Laboratory of Mobile Network and Mobile Multimedia Technology, Shenzhen 518055, China \\
$^{6}$ ZTE Corporation, Shenzhen 518055, China \\
Corresponding author: \textit{Wei Chen, Bo Ai}\\
Email: \{20271232, weich, jialongxu, 23115031, boai\}@bjtu.edu.cn, li.lun1@zte.com.cn
}
}

\maketitle

\begin{abstract}
For massive multiple-input multiple-output systems in the frequency division duplex (FDD) mode, accurate downlink channel state information (CSI) is required at the base station (BS). However, the increasing number of transmit antennas aggravates the feedback overhead of CSI. Recently, deep learning (DL) has shown considerable potential to reduce CSI feedback overhead. In this paper, we propose a Swin Transformer-based autoencoder network called SwinCFNet for the CSI feedback task. In particular, the proposed method can effectively capture the long-range dependence information of CSI. Moreover, we explore the impact of the number of Swin Transformer blocks and the dimension of feature channels on the performance of SwinCFNet. Experimental results show that SwinCFNet significantly outperforms other DL-based methods with comparable model sizes, especially for the outdoor scenario.
\end{abstract}


\section{Introduction}
As one of the key technologies for the fifth generation (5G) wireless communication system, massive multiple-input multiple-output (MIMO) can effectively improve the throughput, enhance the coverage, and promise the reliability of the 5G system by deploying a large number of antennas at both the base station (BS) and the user equipment (UE) \cite{1}. These benefits highly depend on the accurate acquisition of downlink channel state information (CSI) at the BS. For the time division duplex (TDD) mode, the BS estimates the uplink CSI through the pilot signal transmitted by the UE and infers the downlink CSI according to the channel reciprocity \cite{2}. For the frequency division duplex (FDD) mode, channel reciprocity no longer holds due to the different frequency bands assigned for the uplink transmission and the downlink transmission. Therefore, to support the FDD mode, the estimated downlink CSI at the UE side based on pilot signals should be further fed back from the UE to the BS side. However, this CSI feedback mechanism inevitably consumes the uplink resources, and reduces the uplink efficiency. Furthermore, 6G is anticipated to leverage ultra-massive MIMO to further improve transmission data rate, meeting the requirement of emerging applications, e.g., immersive communications \cite{3}. The growth of antennas in ultra-massive MIMO will lead to CSI matrices of a huge size, and aggravate the proportion of uplink resources in FDD mode. It is essential to create an efficient technique to compress the CSI matrix in order to reduce the feedback overhead.

Traditional compressive sensing (CS)-based methods exploit the sparse characteristic of the channel in the transform domain to compress the CSI information \cite{4}. However, in practical communication scenarios, the channel of massive MIMO is not strictly sparse, preventing CS-based methods from achieving optimal performance. Furthermore, iterative algorithms for CSI reconstruction are usually too complex to be applied in practical scenarios.

Recently, deep learning (DL) has shown great potential to revolutionize communication systems \cite{5}, and promote the development of semantic communications \cite{6}. CSI compression and feedback in massive MIMO can also exploit the powerful DL \cite{7,8,9,10,11,12,13,14,15,16,17}. The first DL-based CSI feedback method called CsiNet was proposed in \cite{7}, which was designed based on the structure of the autoencoder. The proposed CsiNet uses an encoder to compress the CSI into codewords at the UE and a decoder to reconstruct the CSI from these codewords at the BS, which outperforms CS-based methods. The attention mechanism is introduced to DL-based CSI feedback by Attention-CsiNet \cite{8} and LSTM-attention CsiNet \cite{9} to fully utilize different feature maps. CRNet \cite{10} uses a multi-resolution architecture in the network and emphasizes the importance of the training scheme. Besides, there are a lot of novel neural networks that focus on improving feedback accuracy, such as DS-NLCsiNet \cite{11}, CLNet \cite{12} and DFECsiNet \cite{13}. Furthermore, deep joint source-channel coding based CSI feedback is demonstrated superior to the separate source-channel coding based schemes \cite{14}. Different from the two-sided framework adopted in the aforementioned methods, a one-sided deep learning framework \cite{15} and a novel plug-and-play method are further proposed for the CSI feedback \cite{16}. However, most of the aforementioned works are based on the convolutional neural network (CNN), which is inefficient in capturing the global dependencies of the CSI. Inspired by the excellent performance of the transformer, a two-layer transformer architecture named TransNet was proposed in \cite{17}, which greatly improves the feedback performance.

In the field of computer vision, vision Transformer (ViT) first applied the transformer to image classification and provided a vision backbone network \cite{18}. To avoid the quadratic complexity of ViT, Swin Transformer was proposed in \cite{19}, which computed self-attention within non-overlapping local windows. Compared with ViT, Swin Transformer extracts features of varying scales more effectively, which enhances the modeling ability. Inspired by its tremendous success, a wireless image transmission transformer has been built  to extract long-term image representation \cite{20}, which significantly outperforms the CNN-based scheme \cite{21}. 

In this paper, we propose a novel autoencoder framework for CSI feedback named SwinCFNet, which exploits the Swin Transformer. The framework contains two stages to construct long-term hierarchical features, and each has an even number of Swin Transformer blocks. The main contributions of this paper are summarized as follows.
\begin{itemize}
\item We design a SwinCFNet network built upon the Swin Transformer to deal with the CSI feedback task. Our proposed architecture uses Swin Transformer blocks to exploit inter-frequency and inter-antenna correlations in the channel matrix. These blocks utilize a shifted window-based multi-head self-attention mechanism to extract long-range information of CSI, which effectively obtains the latent CSI representation.
\item We verify the performance of the proposed SwinCFNet network with different compression ratios. We also investigate the effect of the number of Swin Transformer blocks and the dimension of feature channels on the SwinCFNet. Simulation results demonstrate that our proposed framework can significantly improve the accuracy of the feedback, especially for the outdoor scenario.
\end{itemize}

\section{System model}

\subsection{Massive MIMO-OFDM System}

We consider a single-cell FDD massive MIMO system with $N_{t}$ antennas at the BS and single antenna at the UE. Orthogonal frequency division multiplexing (OFDM) with $\tilde{N}_{c}$ subcarriers is used in this model. The received signal at the $\textit{n}$-th subcarrier can be expressed as
\begin{equation}
y_{n} ={\tilde{\mathbf{h}}_{n}^{H}}\mathbf{v}_{n}x_{n}+z_{n},\label{eq1}
\end{equation}
where $\tilde{\mathbf{h}}_{n}\in \mathbb{C}^{N_{t}}$, $\mathbf{v}_{n}\in \mathbb{C}^{N_{t}}$, $x_{n}\in \mathbb{C}$, $z_{n}\in \mathbb{C}$ denote the channel vector, the precoding vector, the data symbol transmitted in the downlink and the additive noise at the $\textit{n}$-th subcarrier, respectively. The downlink CSI is expressed as
\begin{equation}
\tilde{\mathbf{H}} =[\tilde{\mathbf{h}}_{1},\tilde {\mathbf{h}}_{2},...,\tilde{\mathbf{h}}_{\tilde N_{c}}]^T \in \tilde N_{c}\times N_{t}.\label{eq2}
\end{equation}

The total number of feedback parameters is $N_{t}\tilde N_{c}$, which is proportional to the number of antennas. With the increase of antennas in future massive MIMO systems, the size of CSI matrix may surpass the feedback overhead provided by the uplink. Fortunately, the CSI matrix $\tilde {\mathbf{H}}$ is sparse in the angular-delay domain, which makes it easier to compress \cite{7}. The channel matrix can be transformed from the spatial-frequency domain to the angular-delay domain by a 2D discrete Fourier transform (DFT) expressed as:
\begin{equation}
\mathbf{H}'=\mathbf{F}_{c}\tilde{\mathbf{H}}\mathbf{F}_{t}^{H},\label{eq3}
\end{equation}
where $\mathbf{F}_{c}$ and $\mathbf{F}_{t}$ are DFT matrices with $\tilde N_{c}\times\tilde N_{c}$ and $N_{t}\times N_{t}$ elements, respectively. In the angular-delay domain, only the first $N_{c}$ rows of $\mathbf{H}'$ have significant components. The subsequent rows consist mostly of elements near zero and can be omitted. We retain only the initial $N_{c}$ rows of $\mathbf{H}'$, representing the new CSI matrix with $\mathbf{H}\in \mathbb{C}^{N_{c} \times N_{t}}$. Since the elements of $\mathbf{H}$ are complex numbers, both their real and imaginary parts need to be fed back. Thus, the total number of CSI parameters is reduced to 2$N_{c}N_{t}$.

\subsection{DL-Based CSI Feedback}
In this paper, we design an autoencoder based neural network for CSI compression. The UE encodes the channel matrix $\mathbf{H}$ into a vector $\mathbf{s}$ with the length satisfying $M<2N_{c}N_{t}$, which is expressed as:
\begin{equation}
\mathbf{s}=f_{en}(\mathbf{H}, \Theta_{en}),\label{eq4}
\end{equation}
where $f_{en}(\cdot)$ represents the function of the encoder and $\Theta_{en}$ denotes the corresponding parameters set. The compression ratio is defined as $\gamma=\frac{M}{2N_{c}N_{t}}$. The BS recovers the original channel matrix from the codeword $\mathbf{s}$ as follows:
\begin{equation}
\hat{\mathbf{H}}=f_{de}(\mathbf{s}, \Theta_{de}),\label{eq5}
\end{equation}
where $f_{de}(\cdot)$ represents the function of the decoder and $\Theta_{de}$ denotes the parameter set of the decoder. $\hat{\mathbf{H}}$ is the recovered channel matrix in the angular delay domain. Then the channel matrix in the spatial-frequency domain can be obtained by applying zero filling and inverse DFT.

\begin{figure*}[!t]
\centering 
\includegraphics[width=2.1\columnwidth]{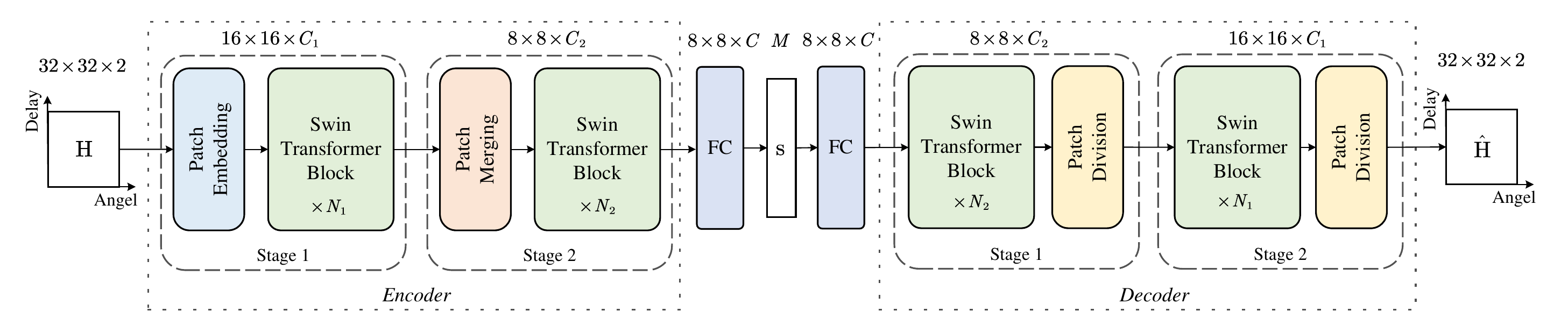}
\caption{The overall architecture of the proposed SwinCFNet, which includes the encoder and the decoder.}
\label{fig}
\end{figure*}

\begin{figure}[!t]
\centerline{\includegraphics[width=0.6\columnwidth]{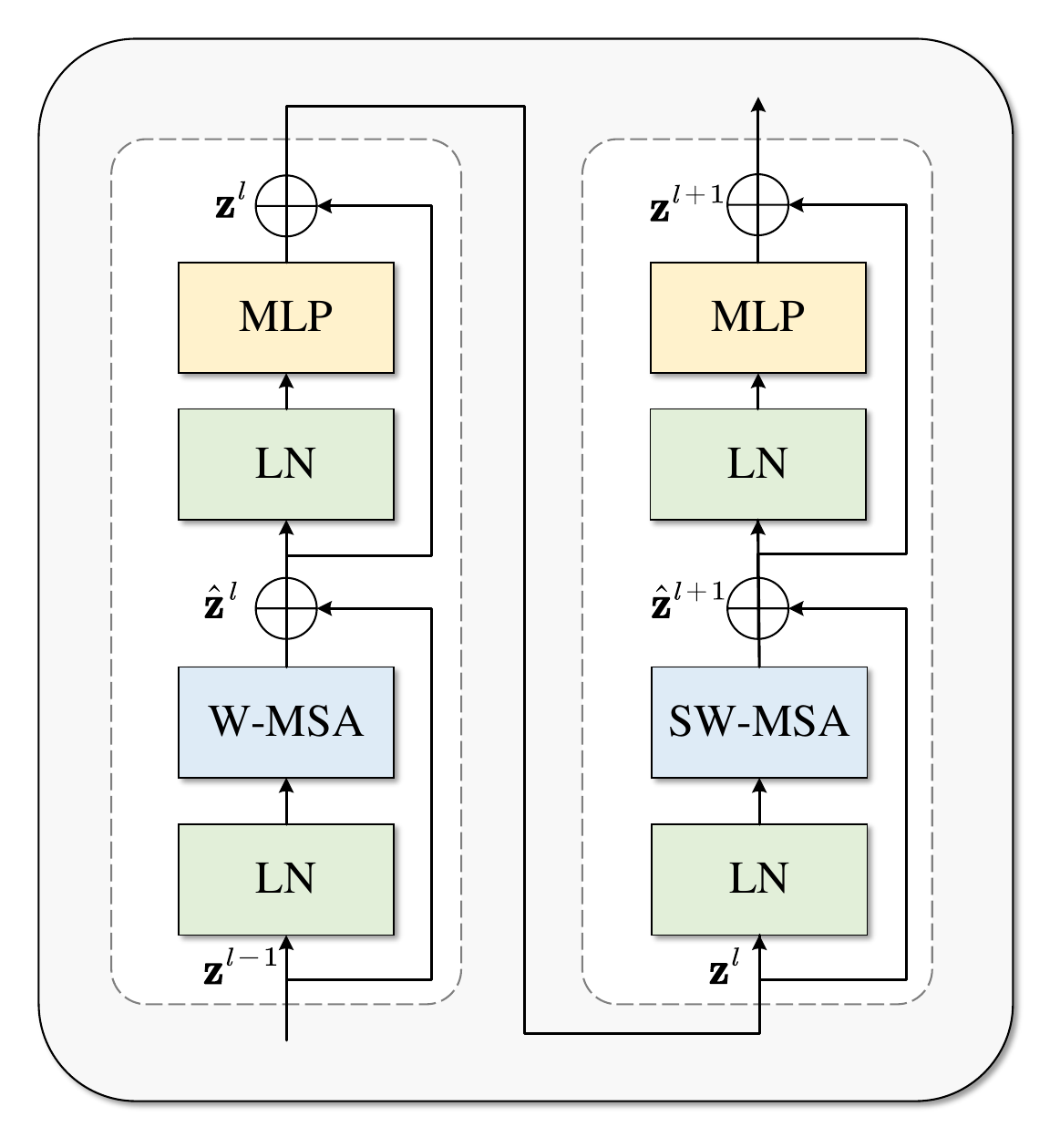}}
\caption{Two successive Swin Transformer Blocks.}
\label{fig2}
\end{figure}

\section{Design of SwinCFNet}

\subsection{Overall Architecture}
Inspired by the Swin Transformer \cite{19}, we construct our CSI feedback network, namely SwinCFNet. An overview of the SwinCFNet architecture is presented in Fig. \ref{fig}. The CSI matrix in the angular-delay domain is treated as an input image of size $N_{c}\times N_{t}\times2$, and the two channels correspond to the real and imaginary parts of the CSI matrix. The patch embedding module consists of patch partition and linear embedding. In our implementation, the patch size is configured to $2\times2$ and each patch is treated as a "token". Firstly, it splits the input CSI image into $\frac{N_{c}}{2}\times \frac{N_{t}}{2}$ non-overlapping patches through a patch splitting module, and the feature dimension of each patch is $2\times2\times2=8$. After the patch partition, a linear embedding layer projects the patched features into the pre-assigned embedding dimension $C_{1}$.

Then, $N_1$ Swin Transformer blocks are applied on these patch tokens, which maintain the number of tokens. These blocks together with the patch embedding module are referred to as ``Stage 1". Subsequently, these patch tokens are fed into a patch merging layer to reduce the feature size as the network gets deeper. Specifically, the patch merging layer concatenates features of each group of $2\times2$ neighboring patches and applies a linear layer on the concatenated features. After that, it feeds the $\frac{N_{c}}{4}\times \frac{N_{t}}{4}$ patch tokens to $N_2$ Swin Transformer blocks for feature transformation. $N_2$ Swin Transformer blocks with the patch merging block are denoted as ``Stage 2". Next, a fully connected (FC) layer is used on the feature to project it to the vector with $M=8\times8\times C$ dimension. Thus, we obtain a compressed feature vector $\mathbf{s}\in \mathbb{R}^{M}$ for feedback.

At the receiver, the SwinCFNet decoder $f_{de}$ has a symmetric architecture with the encoder $f_{en}$. The decoder is composed of an FC layer and two stages. Each stage consists of a pre-assigned number of Swin Transformer blocks and a patch division layer for up-sampling. Finally, the received feature vector $\mathbf{s}$ is reconstructed into $\hat{\mathbf{H}}$ by the decoder.

To train the SwinCFNet, we adopt the end-to-end method. The set of parameters is denoted as $\Theta = \{\Theta_{en}, \Theta_{de}\}$, which is updated by the adaptive moment estimation (ADAM) optimizer. The training loss function of our network is the mean square error (MSE), which can be calculated as follows:
\begin{equation}
L(\Theta)=\frac{1}{T} \sum_{i=1}^{T} \left \|f(\mathbf{s}_i;\Theta)-\mathbf{H}_i\right \|_2^2 \label{eq6}
\end{equation}
where $T$ is the total number of samples in the training set, the subscript denotes the $i$-th sample in the training set, and norm $\left \| \cdot  \right \| _2$ denotes the Euclidean norm.

\subsection{Swin Transformer Blocks}
In the standard multi-head self-attention (MSA) mechanism, each token needs to be computed based on its relationship to all other tokens, where the computational complexity is quadratic to the number of tokens. As illustrated in Fig. \ref{fig2}, Swin Transformer \cite{19} proposes a window-based MSA (W-MSA) followed by a shifted window-based MSA (SW-MSA) module, which has linear computational complexity to image size. A Swin Transformer block consists of two consecutive modules built up by a SW-MSA module and a 2-layer multilayer perceptron (MLP) with a LayerNorm (LN) layer and residual connection. MLP consists of a hidden layer and an embedding layer with the GELU activation. The SW-MSA mechanism can fully pay attention to the correlation characteristics between different antennas and subcarriers while taking into account the long-range dependency.

\begin{figure}[!t]
    \centering
    \subfloat[]{
        \includegraphics[width=0.8\columnwidth]{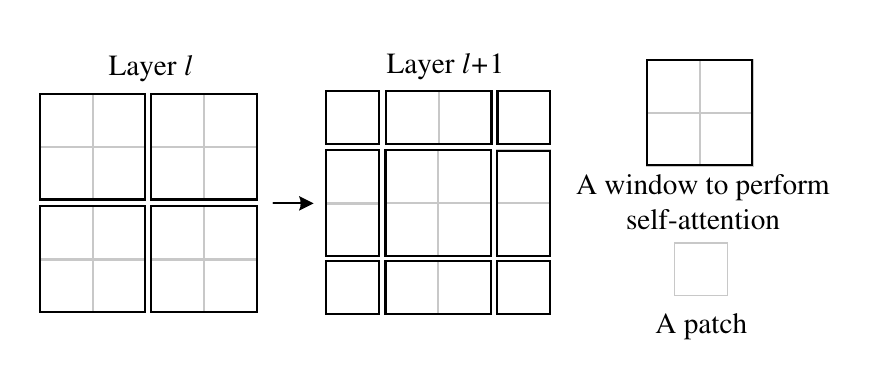}
        }\\
    \subfloat[]{
        \includegraphics[width=1.0\columnwidth]{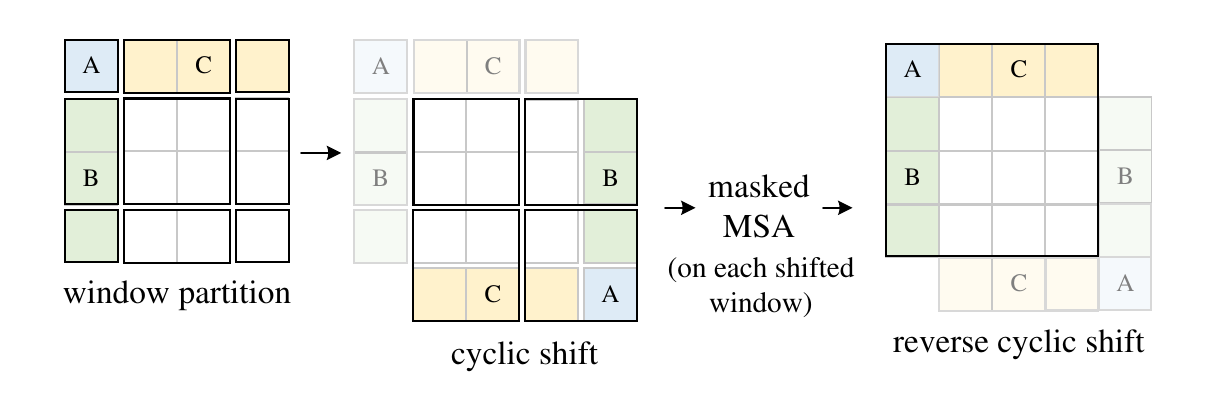}
        }
    \caption{(a) Shifted window approach for computing self-attention. (b) Implementation of the shifted window partition with the cyclic shift.}
    \label{fig3}
\end{figure}

As shown in Fig. \ref{fig3} (a), Swin Transformer blocks divide the input into many small patches. In our implementation, the window size is set to 2 and thus each independent window consists of $2\times2$ patches. The regular window partitioning is adopted in layer $l$, and it computes self-attention separately for each window to reduce computation overhead. In the next layer $l+1$, the window partition is shifted to enable the interaction of adjacent pixels in different windows. Therefore, the self-attention computed in the new windows crosses the boundaries of the previous windows, thereby establishing connections among them. As illustrated in Fig. \ref{fig3} (b), a cyclic-shifting mechanism is employed for the shifted window partition to gain the same windows as the regular partition. After the cyclic shift, the new window may consist of several non-adjacent sub-windows in the original feature map. In order to prevent interference among these non-adjacent sub-windows, a masking approach is applied to limit self-attention computation to each sub-window. 

Given a window feature $z^{l}$, the $query$, $key$ and $value$ matrices $Q$, $K$ and $V$ are computed as
\begin{equation}
\begin{aligned}
&Q=z^{l}W_{Q},\,K=z^{l}W_{K},\,V=z^{l}W_{V},\label{eq7}
\end{aligned}
\end{equation}
where $W_Q$, $W_K$, $W_V$ are projection matrices shared across different windows. Then, the self-attention computed in W-MSA and SW-MSA can be calculated as follows
\begin{equation}
\begin{aligned}
&\mbox{Attention}(Q,K,V)=\mbox{SoftMax}(QK^{T}/\sqrt{d}+B)V,\label{eq8}
\end{aligned}
\end{equation}
where $d$ represents the dimension of $query$ or $key$. $B$ denotes the learnable relative position bias. With the shifted window partitioning mechanism, consecutive Swin Transformer blocks are computed as
\begin{equation}
\begin{aligned}
&\hat{\mathbf{z}}^{l}=\mbox{W-MSA}\,(\mbox{LN}\,(\mathbf{z}^{l-1}))+\mathbf{z}^{l-1}, \\
&\mathbf{z}^{l}=\mbox{MLP}\,(\mbox{LN}\,(\hat{\mathbf{z}}^{l}))+\hat{\mathbf{z}}^{l}, \\
&\hat{\mathbf{z}}^{l+1}=\mbox{SW-MSA}\,(\mbox{LN}\,(\mathbf{z}^{l}))+\mathbf{z}^{l}, \\
&\mathbf{z}^{l+1}=\mbox{MLP}\,(\mbox{LN}\,(\hat{\mathbf{z}}^{l+1}))+\hat{\mathbf{z}}^{l+1}, \label{eq9}
\end{aligned}
\end{equation}
where $\hat{\mathbf{z}}^{l}$ and $\mathbf{z}^{l}$ stand for the output features of the (S)W-MSA module and the MLP module for block $\mathit{l}$, respectively. 

\section{Experimental Results}
In this section, we introduce the experimental settings, compare the performance of the proposed SwinCFNet network with existing DL-based feedback methods, and investigate the effect of the number of Swin Transformer blocks and the dimension of feature channels on the SwinCFNet.

\subsection{Experimental Settings}
In order to fairly compare our experimental results with others, we use the dataset generated by COST2100 \cite{22}. Two types of scenarios are considered, i.e., the indoor scenario at the 5.3 GHz band and outdoor scenario at the 300 MHz band. We set $\tilde N_{c}$ = 1024 subcarriers and $N_{t}$ = 32 uniform linear array antennas at the BS. After the angle delay domain transformation, the first 32 rows of $\mathbf{H}'$ are reserved due to the sparsity of the CSI matrix and the final CSI is $\mathbf{H}\in \mathbb{C}^{{32} \times{32}}$. The training, validation and testing sets contain 100,000, 30,000, and 20,000 samples, respectively. 

For both the indoor and outdoor scenarios, the window size of the proposed method is set to 4 and the patch size is set to $2\times2$. In addition, multi-head self-attention within local or shifted windows is used to reduce the computation complexity. The number of heads for each layer is set to $[4,8]$ and the expansion layer of each MLP is set to 4 for all experiments. The number of Swin Transformer blocks $[N_1, N_2]$ is set to $[2,4]$. In the indoor scenario, the CSI matrix has few non-zero elements. Therefore, the dimension of the feature channel $[C_1, C_2]$ is set to $[64,128]$, which achieves good performance. However, in the outdoor scenario, the non-zero points become scattered and the structure of the CSI matrix is more complex. In general, more features always require a larger network to enrich the expressive ability. Therefore, the dimension of the feature channel $[C_1, C_2]$ is set to $[128,256]$ in the outdoor scenario to get better performance. In addition, the size of the feature dimension $C$ determines the compression rate of the model. 

We train the SwinCFNet with the batch size of 200 on a single NVIDIA GTX 1080Ti GPU and update the parameters with a constant learning rate of $1\times10^{-4}$. The largest training epoch is set as 1000. The performance of the network architecture is evaluated by normalized mean square error (NMSE) and cosine similarity. NMSE between the original $\mathbf{H}$ and the reconstructed $\hat{\mathbf{H}}$ is used to measure the recovery performance, which is given by
\begin{equation}
NMSE=E\left\{ \left\| \mathbf{H}-\hat{\mathbf{H}} \right\|_{2}^{2}/ \left\|\mathbf{H}\right\|_{2}^{2} \right\}.\label{eq10}
\end{equation}
The cosine similarity computed in the spatial-frequency domain is also used to compare the performance of different methods, which is given by
\begin{equation}
\rho=E\left\{\frac{1}{\tilde{N_{c}}}\sum_{i=1}^ {\tilde{N_{c}}}\frac{\left| {\hat{\tilde{\mathbf{h}}}}_{i}^{H} \tilde{\mathbf{h}}_{i}\right|}{\left\| {\hat{\tilde{\mathbf{h}}}}_{i}\right\|_{2} \left\| \tilde{\mathbf{h}}_{i}\right\|_{2}} \right\},\label{eq11}
\end{equation}
where $\tilde{\mathbf{h}}_{i}$ and ${\hat{\tilde{\mathbf{h}}}}_{i}$ denote the original and reconstructed channel vector of the $\mathit{n}$-th subcarrier, respectively.

\subsection{Performance Comparison}

\begin{table*}[!t]
\renewcommand{\arraystretch}{1.25}
\centering
\caption{NMSE(dB) and $\rho$ Performance of the NNs Using the Dataset Generated by COST2100}
\resizebox{\textwidth}{!}{%
\begin{tabular}{c|cccc|cccc|cccc|cccc}
\hline
CR &
  \multicolumn{4}{c|}{1/4} &
  \multicolumn{4}{c|}{1/8} &
  \multicolumn{4}{c|}{1/16} &
  \multicolumn{4}{c}{1/32} \\ \hline
Scenarios &
  \multicolumn{2}{c|}{Indoor} &
  \multicolumn{2}{c|}{Outdoor} &
  \multicolumn{2}{c|}{Indoor} &
  \multicolumn{2}{c|}{Outdoor} &
  \multicolumn{2}{c|}{Indoor} &
  \multicolumn{2}{c|}{Outdoor} &
  \multicolumn{2}{c|}{Indoor} &
  \multicolumn{2}{c}{Outdoor} \\ \hline
Performance &
  \multicolumn{1}{c|}{NMSE} &
  \multicolumn{1}{c|}{$\rho$} &
  \multicolumn{1}{c|}{NMSE} &
  $\rho$ &
  \multicolumn{1}{c|}{NMSE} &
  \multicolumn{1}{c|}{$\rho$} &
  \multicolumn{1}{c|}{NMSE} &
  $\rho$ &
  \multicolumn{1}{c|}{NMSE} &
  \multicolumn{1}{c|}{$\rho$} &
  \multicolumn{1}{c|}{NMSE} &
  $\rho$ &
  \multicolumn{1}{c|}{NMSE} &
  \multicolumn{1}{c|}{$\rho$} &
  \multicolumn{1}{c|}{NMSE} &
  $\rho$ \\ \hline
CsiNet \cite{7} &
  \multicolumn{1}{c|}{-17.36} &
  \multicolumn{1}{c|}{0.99} &
  \multicolumn{1}{c|}{-8.75} &
  0.91 &
  \multicolumn{1}{c|}{\textbackslash{}} &
  \multicolumn{1}{c|}{\textbackslash{}} &
  \multicolumn{1}{c|}{\textbackslash{}} &
  \textbackslash{} &
  \multicolumn{1}{c|}{-8.65} &
  \multicolumn{1}{c|}{0.93} &
  \multicolumn{1}{c|}{-4.51} &
  0.79 &
  \multicolumn{1}{c|}{-6.24} &
  \multicolumn{1}{c|}{0.89} &
  \multicolumn{1}{c|}{-2.81} &
  0.67 \\
Attention-CsiNet \cite{8} &
  \multicolumn{1}{c|}{-20.29} &
  \multicolumn{1}{c|}{0.99} &
  \multicolumn{1}{c|}{-10.43} &
  0.94 &
  \multicolumn{1}{c|}{\textbackslash{}} &
  \multicolumn{1}{c|}{\textbackslash{}} &
  \multicolumn{1}{c|}{\textbackslash{}} &
  \textbackslash{} &
  \multicolumn{1}{c|}{-10.16} &
  \multicolumn{1}{c|}{0.95} &
  \multicolumn{1}{c|}{-6.11} &
  0.85 &
  \multicolumn{1}{c|}{-8.58} &
  \multicolumn{1}{c|}{0.93} &
  \multicolumn{1}{c|}{-4.57} &
  0.79 \\
LSTM-Attention CsiNet \cite{9} &
  \multicolumn{1}{c|}{-22.00} &
  \multicolumn{1}{c|}{0.99} &
  \multicolumn{1}{c|}{-10.20} &
  0.93 &
  \multicolumn{1}{c|}{\textbackslash{}} &
  \multicolumn{1}{c|}{\textbackslash{}} &
  \multicolumn{1}{c|}{\textbackslash{}} &
  \textbackslash{} &
  \multicolumn{1}{c|}{-11.00} &
  \multicolumn{1}{c|}{0.96} &
  \multicolumn{1}{c|}{-5.80} &
  0.82 &
  \multicolumn{1}{c|}{-8.80} &
  \multicolumn{1}{c|}{0.93} &
  \multicolumn{1}{c|}{-3.70} &
  0.71 \\
CRNet \cite{10} &
  \multicolumn{1}{c|}{-26.99} &
  \multicolumn{1}{c|}{\textbackslash{}} &
  \multicolumn{1}{c|}{-12.71} &
  \textbackslash{} &
  \multicolumn{1}{c|}{-16.01} &
  \multicolumn{1}{c|}{\textbackslash{}} &
  \multicolumn{1}{c|}{-8.04} &
  \textbackslash{} &
  \multicolumn{1}{c|}{-11.35} &
  \multicolumn{1}{c|}{\textbackslash{}} &
  \multicolumn{1}{c|}{-5.44} &
  \textbackslash{} &
  \multicolumn{1}{c|}{-8.93} &
  \multicolumn{1}{c|}{\textbackslash{}} &
  \multicolumn{1}{c|}{-3.51} &
  \textbackslash{} \\
DS-NLCsiNet \cite{11} &
  \multicolumn{1}{c|}{-24.99} &
  \multicolumn{1}{c|}{0.99} &
  \multicolumn{1}{c|}{-12.09} &
  0.95 &
  \multicolumn{1}{c|}{-17.00} &
  \multicolumn{1}{c|}{0.99} &
  \multicolumn{1}{c|}{-7.96} &
  0.90 &
  \multicolumn{1}{c|}{-12.93} &
  \multicolumn{1}{c|}{0.97} &
  \multicolumn{1}{c|}{-4.98} &
  0.81 &
  \multicolumn{1}{c|}{-8.64} &
  \multicolumn{1}{c|}{0.93} &
  \multicolumn{1}{c|}{-3.35} &
  0.73 \\
CLNet \cite{12} &
  \multicolumn{1}{c|}{-29.16} &
  \multicolumn{1}{c|}{\textbackslash{}} &
  \multicolumn{1}{c|}{-12.88} &
  \textbackslash{} &
  \multicolumn{1}{c|}{-15.60} &
  \multicolumn{1}{c|}{\textbackslash{}} &
  \multicolumn{1}{c|}{-8.29} &
  \textbackslash{} &
  \multicolumn{1}{c|}{-11.15} &
  \multicolumn{1}{c|}{\textbackslash{}} &
  \multicolumn{1}{c|}{-5.56} &
  \textbackslash{} &
  \multicolumn{1}{c|}{-8.95} &
  \multicolumn{1}{c|}{\textbackslash{}} &
  \multicolumn{1}{c|}{-3.49} &
  \textbackslash{} \\
DFECsiNet \cite{13} &
  \multicolumn{1}{c|}{-27.51} &
  \multicolumn{1}{c|}{0.99} &
  \multicolumn{1}{c|}{-12.30} &
  0.95 &
  \multicolumn{1}{c|}{-16.36} &
  \multicolumn{1}{c|}{0.98} &
  \multicolumn{1}{c|}{-7.88} &
  0.89 &
  \multicolumn{1}{c|}{-12.09} &
  \multicolumn{1}{c|}{0.96} &
  \multicolumn{1}{c|}{-5.34} &
  0.82 &
  \multicolumn{1}{c|}{-9.10} &
  \multicolumn{1}{c|}{0.93} &
  \multicolumn{1}{c|}{-3.38} &
  0.73 \\
TransNet \cite{15} &
  \multicolumn{1}{c|}{\textbf{-32.38}} &
  \multicolumn{1}{c|}{\textbackslash{}} &
  \multicolumn{1}{c|}{-14.86} &
  \textbackslash{} &
  \multicolumn{1}{c|}{\textbf{-22.91}} &
  \multicolumn{1}{c|}{\textbackslash{}} &
  \multicolumn{1}{c|}{-9.99} &
  \textbackslash{} &
  \multicolumn{1}{c|}{-15.00} &
  \multicolumn{1}{c|}{\textbackslash{}} &
  \multicolumn{1}{c|}{-7.82} &
  \textbackslash{} &
  \multicolumn{1}{c|}{\textbf{-10.49}} &
  \multicolumn{1}{c|}{\textbackslash{}} &
  \multicolumn{1}{c|}{-4.13} &
  \textbackslash{} \\
\textbf{SwinCFNet(Ours)} &
  \multicolumn{1}{c|}{-28.94} &
  \multicolumn{1}{c|}{\textbf{0.99}} &
  \multicolumn{1}{c|}{\textbf{-21.29}} &
  \textbf{0.98} &
  \multicolumn{1}{c|}{-21.09} &
  \multicolumn{1}{c|}{\textbf{0.99}} &
  \multicolumn{1}{c|}{\textbf{-15.90}} &
  \textbf{0.97} &
  \multicolumn{1}{c|}{\textbf{-15.12}} &
  \multicolumn{1}{c|}{\textbf{0.98}} &
  \multicolumn{1}{c|}{\textbf{-11.50}} &
  \textbf{0.94} &
  \multicolumn{1}{c|}{-9.39} &
  \multicolumn{1}{c|}{\textbf{0.94}} &
  \multicolumn{1}{c|}{\textbf{-7.62}} &
  \textbf{0.89} \\ \hline
\multicolumn{17}{l}{$^{\mathrm{a}}$ ``\textbackslash{}'' means the performance is not reported in the original paper.}
\end{tabular}%
}
\label{tab1}
\end{table*}

The performance of NMSE and $\rho$ are summarized in Table \ref{tab1}, where the compression rate (CR) is set to 1/4, 1/8, 1/16 and 1/32. The best results of all networks are shown in bold. In the indoor scenario, the NMSE values of most existing methods are below -20 dB when the compression ratio is 1/4, which achieves good CSI accuracy. However, in the outdoor scenario, the NMSE values of the existing DL-based methods are far from -20 dB with CR = 1/4, while SwinCFNet achieves an impressive NMSE value of -21.29 dB. It is worth noting that our proposed method provides significant performance improvement in outdoor scenario. 

\begin{figure}[!t]
\centerline{\includegraphics[width=0.85\columnwidth]{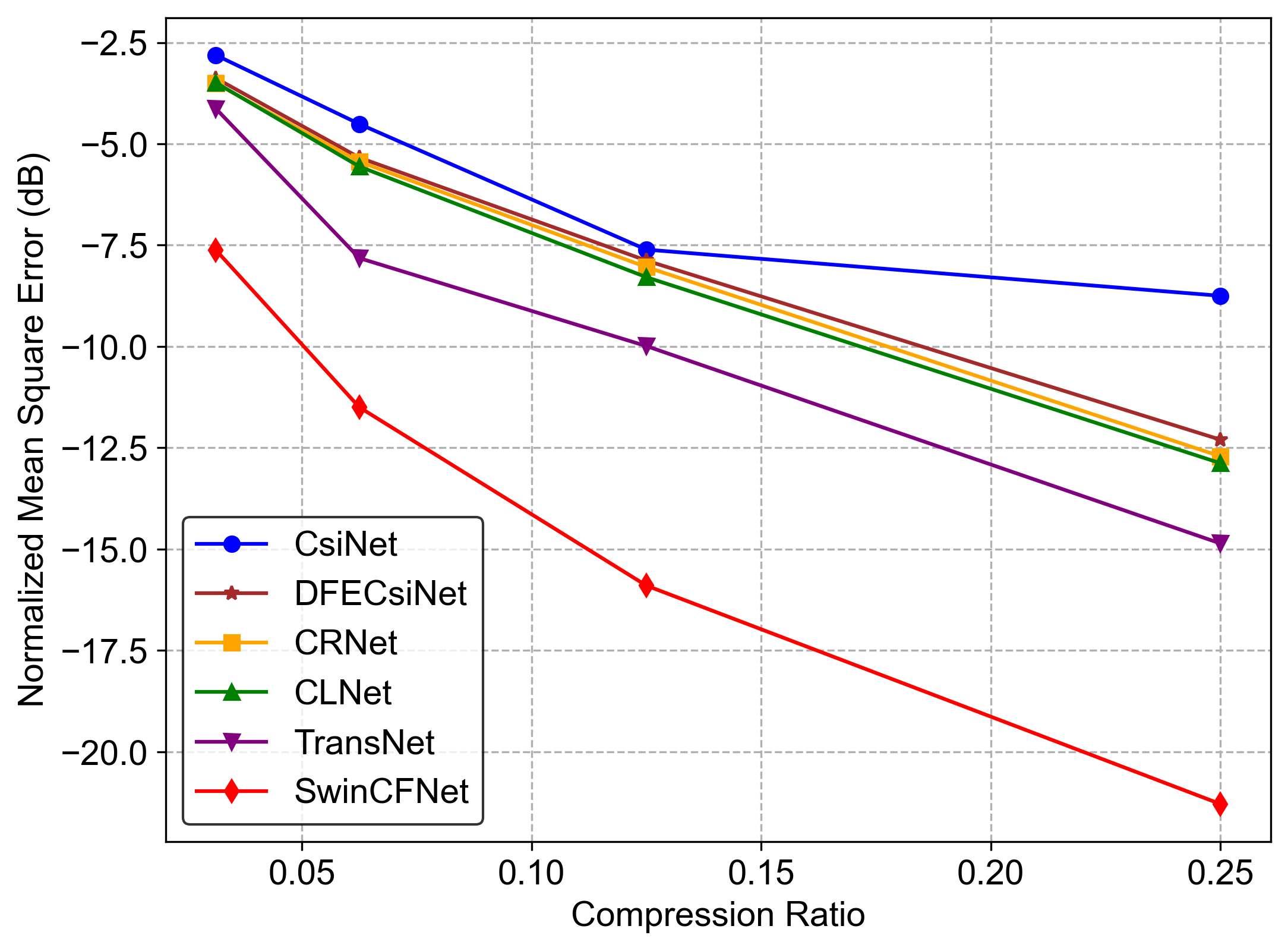}}
\caption{NMSE(dB) performance comparison under different compression ratios in the outdoor scenario.}
\label{fig4}
\end{figure}

From Table \ref{tab1}, we can conclude that our proposed framework is able to achieve the lowest NMSE values and the highest $\rho$ values at all compression ratios for the outdoor scenario. Compared to the best of the other methods, SwinCFNet improves accuracy by 77.25\% and 74.36\% in the outdoor scenario with CR = 1/4, 1/8, respectively. SwinCFNet also performs well at CR = 1/16, 1/32, achieving 57.15\% and 50.10\% gains in accuracy than the best results of other methods, respectively. Fig. \ref{fig4} intuitively shows the NMSE performance of SwinCFNet and other DL-based networks under different CRs in the outdoor scenario. Limited by the model capacity, it can be observed that traditional methods show poor performance in the outdoor scenario. From Fig. \ref{fig4}, we can see that SwinCFNet outperforms other methods at all CRs. Especially when CR is equal to 1/4 and 1/8, SwinCFNet obtains NMSE gains of more than 5 dB compared to the best of the other methods, which shows a huge performance boost. The main reason is that the hierarchical feature extraction and the shifted window approach make SwinCFNet effectively model long-range dependencies in the channel matrix.

For the indoor scenario, transformer-based models demonstrate superior performance. SwinCFNet has only a small gap in NMSE performance with the advanced transformer-based model TransNet, and the model size of SwinCFNet is smaller. In terms of cosine similarity, SwinCFNet performs better than most other networks in most cases, which is also observed in the NMSE performance. Especially when CR is equal to 1/16, SwinCFNet achieves the best performance among all the DL-based methods listed in Table \ref{tab1} for the indoor scenario. Specifically, SwinCFNet gets a 77.42$\%$ gain in accuracy than CsiNet and obtains NMSE gains of 0.12 dB compared to the best results of other methods in this case. 

The difference in models between the indoor scenario and the outdoor scenario is the depth of the network. The CSI matrix of the indoor scenario is relatively sparse, so larger models slightly improve the performance. For the outdoor CSI matrix, the feature points are more complex and scattered, using a deeper network can increase the representation capacity of the model and achieve better performance. Our proposed SwinCFNet constructs hierarchical feature maps by merging image patches in deeper layers, which has proven beneficial in improving reconstruction performance.

\subsection{Performance of Different Hyper-Parameters Settings}
In order to investigate the effect of hyper-parameters on SwinCFNet, we change the dimension of the feature channel $[C_1, C_2]$ and the number of Swin Transformer blocks $[N_1, N_2]$. We choose the outdoor scenario with CR = 1/4 and provide the complexity analysis of the SwinCFNet with CsiNet and TransNet. The complexity of the network is evaluated by two metrics: the number of network parameters and the number of floating point operations (FLOPs). Simulation results are given in Table \ref{tab2} and Table \ref{tab3}.

SwinCFNet has larger computational complexity (FLOPs) than other DL-based CSI feedback methods. However, the core issue of CSI feedback is still the reconstruction accuracy of CSI, and the number of FLOPs is far from being an obstacle for practical application \cite{17}.

Table \ref{tab2} lists the performance of NMSE, FLOPs, and model size (Params) for CsiNet, TransNet, and SwinCFNet in the outdoor scenario with CR = 1/4. From Table \ref{tab2}, we can find that when $[C_1, C_2]=[64, 128]$, the proposed SwinCFNet has the fewest parameters, which can still achieve better performance than CsiNet and TransNet. In this case, SwinCFNet achieves 116.44$\%$ of the performance of TransNet with 78.48$\%$ parameters. As the value of $[C_1, C_2]$ increases, the performance of the model continues to improve, while the computational complexity also increases. When $[C_1, C_2]=[128, 256]$, the largest SwinCFNet model obtains NMSE gains of 6.43 dB compared to TransNet. On the basis of fully excavating the power of Swin Transformer blocks, the feedback performance is greatly improved. We can conclude that the larger feature channel dimension has a stronger capability to extract CSI features for the outdoor scenario. Furthermore, the proposed network with smaller feature channel dimension still outperforms other DL-based methods with comparable model sizes. 

We further investigate the effect of the number of Swin Transformer blocks $[N_1, N_2]$ on SwinCFNet. Since successive Swin Transformer blocks include a W-MSA module and a SW-MSA module, the number of Swin Transformer blocks is even in each stage. As shown in Table \ref{tab3}, we can conclude that our proposed framework can achieve better NMSE values when $[N_1, N_2]=[2, 4]$. When $[N_1, N_2]=[4, 2]$, the network has similar FLOPs, but the performance is slightly degraded. Therefore, we recommend further study to use $[N_1, N_2] = [2, 4]$ for better feedback performance.
\begin{table}[!t]
\scriptsize
\caption{Performance of Different Methods in Outdoor Scenario \\When CR EQUALS 1/4 }
\centering
\renewcommand{\arraystretch}{1.2}
\resizebox{0.95\columnwidth}{!}{%
\begin{tabular}{c|ccc}
\hline
Methods  & NMSE(dB) & Params(M) & FLOPs(M) \\ \hline
CsiNet                    & -8.75    & 2.10   & 5.42     \\
TransNet                  & -14.86   & 2.37   & 35.72   \\
SwinCFNet - {[}64,128{]}  & -15.94   & 1.86   & 156.92   \\
SwinCFNet - {[}128,128{]} & -16.29   & 2.52   & 313.03   \\
SwinCFNet - {[}128,192{]} & -18.50   & 4.56   & 443.41   \\
SwinCFNet - {[}128,256{]} & -21.29   & 7.38   & 624.12   \\ \hline
\multicolumn{4}{l}{$^{1}$ [64,128] represents the value of $[C_1,C_2]$, the same applies to others.} \\
\multicolumn{4}{l}{$^{2}$ The value of $[N_1,N_2]$ is set to [2,4] in this table.}
\end{tabular}%
}
\label{tab2}
\end{table}

\begin{table}[!t]
\tiny
\scriptsize
\caption{Performance of SwinCFNet with Different Numbers of Swin Transformer Blocks in Outdoor Scenario \\ When CR EQUALS 1/4 }
\centering
\renewcommand{\arraystretch}{1.2}
\tiny
\resizebox{0.85\columnwidth}{!}{%
\begin{tabular}{c|ccc}
\hline
$[N_1, N_2]$  & NMSE(dB) & Params(M) & FLOPs(M) \\ \hline
{[}2,2{]}  & -18.14   & 4.22   & 422.25   \\
{[}4,2{]} & -19.05   & 5.02   & 624.67   \\
{[}2,4{]} & -21.29   & 7.38   & 624.12   \\ \hline
\multicolumn{4}{l}{$^{1}$ The value of $[C_1,C_2]$ is set to [128,256] in this table.} \\ 
\end{tabular}%
}
\label{tab3}
\end{table}

\section{Conclusion}
In this work, we have proposed a Swin Transformer-based neural network called SwinCFNet for CSI feedback. SwinCFNet can capture the long-range dependence information of CSI through a two-stage autoencoder network, which is critical to utilize inherent correlations in the channel matrix. Moreover, the effect of hyper-parameters on the SwinCFNet and the complexity analysis are presented. Experimental results have demonstrated that our proposed SwinCFNet greatly improves the reconstruction quality, especially for the outdoor scenario. For the indoor scenario, our proposed network also achieves superior performance with a smaller model size. 

\bibliographystyle{IEEEtran}
\bibliography{IEEEabrv,myref}

\begin{thebibliography}{10}
\providecommand{\url}[1]{#1}
\csname url@samestyle\endcsname
\providecommand{\newblock}{\relax}
\providecommand{\bibinfo}[2]{#2}
\providecommand{\BIBentrySTDinterwordspacing}{\spaceskip=0pt\relax}
\providecommand{\BIBentryALTinterwordstretchfactor}{4}
\providecommand{\BIBentryALTinterwordspacing}{\spaceskip=\fontdimen2\font plus
\BIBentryALTinterwordstretchfactor\fontdimen3\font minus \fontdimen4\font\relax}
\providecommand{\BIBforeignlanguage}[2]{{%
\expandafter\ifx\csname l@#1\endcsname\relax
\typeout{** WARNING: IEEEtran.bst: No hyphenation pattern has been}%
\typeout{** loaded for the language `#1'. Using the pattern for}%
\typeout{** the default language instead.}%
\else
\language=\csname l@#1\endcsname
\fi
#2}}
\providecommand{\BIBdecl}{\relax}
\BIBdecl

\bibitem{1}
L.~Lu, G.~Y. Li, A.~L. Swindlehurst, A.~Ashikhmin, and R.~Zhang, ``{An overview of massive MIMO: Benefits and challenges},'' \emph{IEEE J. Sel. Topics Signal Process.}, vol.~8, no.~5, pp. 742--758, Jun. 2014.

\bibitem{2}
Y.~Ma, Z.~Yuan, W.~Li, and Z.~Li, ``{Truly grant-free technologies and protocols for 6G},'' \emph{ZTE Commun.}, vol.~19, no.~4, pp. 105--110, Dec. 2022.

\bibitem{3}
F.~Zhu, ``{Next generation semantic and spatial joint perception—neural metric-semantic understanding},'' \emph{ZTE Commun.}, vol.~19, no.~1, pp. 61--71, Mar. 2021.

\bibitem{4}
Z.~Qin, J.~Fan, Y.~Liu, Y.~Gao, and G.~Y. Li, ``{Sparse representation for wireless communications: A compressive sensing approach},'' \emph{IEEE Signal Process. Mag.}, vol.~35, no.~3, pp. 40--58, May 2018.

\bibitem{5}
Z.~Qin, H.~Ye, G.~Y. Li, and B.-H.~F. Juang, ``{Deep learning in physical layer communications},'' \emph{IEEE Wireless Commun.}, vol.~26, no.~2, pp. 93--99, Mar. 2019.

\bibitem{6}
J.~Xu, T.-Y. Tung, B.~Ai, W.~Chen, Y.~Sun, and D.~G{\"u}nd{\"u}z, ``{Deep Joint Source-Channel Coding for Semantic Communications},'' \emph{IEEE Wireless Commun.}, 2023.

\bibitem{7}
C.-K. Wen, W.-T. Shih, and S.~Jin, ``{Deep learning for massive MIMO CSI feedback},'' \emph{IEEE Wireless Commun. Lett.}, vol.~7, no.~5, pp. 748--751, Oct. 2018.

\bibitem{8}
Q.~Cai, C.~Dong, and K.~Niu, ``{Attention model for massive MIMO CSI compression feedback and recovery},'' in \emph{Proc. {IEEE} Wireless Commun. Netw. Conf. (WCNC)}, Apr. 2019, pp. 1--5.

\bibitem{9}
Q.~Li, A.~Zhang, P.~Liu, J.~Li, and C.~Li, ``{A novel CSI feedback approach for massive MIMO using LSTM-attention CNN},'' \emph{IEEE Access}, vol.~8, pp. 7295--7302, 2020.

\bibitem{10}
Z.~Lu, J.~Wang, and J.~Song, ``{Multi-resolution CSI feedback with deep learning in massive MIMO system},'' in \emph{Proc. {IEEE} Int. Conf. Commun. (ICC)}, Jun. 2020, pp. 1--6.

\bibitem{11}
X.~Yu, X.~Li, H.~Wu, and Y.~Bai, ``{DS-NLCsiNet: Exploiting non-local neural networks for massive MIMO CSI feedback},'' \emph{IEEE Commun. Lett.}, vol.~24, no.~12, pp. 2790--2794, Dec. 2020.

\bibitem{12}
S.~Ji and M.~Li, ``{CLNet: Complex input lightweight neural network designed for massive MIMO CSI feedback},'' \emph{IEEE Wireless Commun. Lett.}, vol.~10, no.~10, pp. 2318--2322, Oct. 2021.

\bibitem{13}
Y.~Xu, M.~Zhao, S.~Zhang, and H.~Jin, ``{DFECsiNet: Exploiting diverse channel features for massive MIMO CSI feedback},'' in \emph{Proc. 13th Int. Conf. Wireless Commun. Signal Process. (WCSP)}, Oct. 2021, pp. 1--5.

\bibitem{14}
J.~Xu, B.~Ai, N.~Wang, and W.~Chen, ``{Deep joint source-channel coding for CSI feedback: An end-to-end approach},'' \emph{IEEE J. Sel. Areas Commun.}, vol.~41, no.~1, pp. 260--273, Nov. 2022.

\bibitem{15}
W.~Chen, W.~Wan, S.~Wang, P.~Sun, and B.~Ai, ``{CSI-PPPNet: A One-Sided Deep Learning Framework for Massive MIMO CSI Feedback},'' \emph{arXiv preprint arXiv:2211.15851}, 2022.

\bibitem{16}
W.~Wan, W.~Chen, S.~Wang, G.~Y. Li, and B.~Ai, ``{Deep Plug-and-Play Prior for Massive MIMO Systems},'' \emph{arXiv preprint arXiv:2308.04728}, 2023.

\bibitem{17}
Y.~Cui, A.~Guo, and C.~Song, ``{TransNet: Full attention network for CSI feedback in FDD massive MIMO system},'' \emph{IEEE Wireless Commun. Lett.}, vol.~11, no.~5, pp. 903--907, May 2022.

\bibitem{18}
A.~Dosovitskiy \emph{et~al.}, ``An image is worth 16x16 words: Transformers for image recognition at scale,'' \emph{arXiv:2010.11929}, 2020.

\bibitem{19}
Z.~Liu \emph{et~al.}, ``{Swin transformer: Hierarchical vision transformer using shifted windows},'' in \emph{Proc. IEEE/CVF Int. Conf. Comput. Vis. (ICCV)}, Oct. 2021, pp. 10\,012--10\,022.

\bibitem{20}
K.~Yang, S.~Wang, J.~Dai, K.~Tan, K.~Niu, and P.~Zhang, ``{WITT: A wireless image transmission transformer for semantic communications},'' in \emph{Proc. IEEE Int. Conf. Acoust., Speech Signal Process. (ICASSP)}, Jun. 2023, pp. 1--5.

\bibitem{21}
J.~Xu, B.~Ai, W.~Chen, A.~Yang, P.~Sun, and M.~Rodrigues, ``{Wireless image transmission using deep source channel coding with attention modules},'' \emph{IEEE Trans. Circuits Syst. Video Technol.}, vol.~32, no.~4, pp. 2315--2328, Apr. 2021.

\bibitem{22}
L.~Liu \emph{et~al.}, ``{The COST 2100 MIMO channel model},'' \emph{IEEE Wireless Commun.}, vol.~19, no.~6, pp. 92--99, Dec. 2012.

\end{thebibliography}

\end{document}